\documentstyle[amssymb,epsfig]{mn}

\title[FR-II radio sources and the ICM]{On the interaction of FR-II 
radio sources with the intracluster medium}
\author[P. Alexander]
       {P. Alexander\\
        Astrophysics Group, Cavendish Laboratory, Madingley Road, Cambridge CB3 0HE}
\date{Accepted ??????.
      Received ??????;
      in original form ??????}

\pagerange{\pageref{firstpage}--\pageref{lastpage}}
\pubyear{2002}

\begin{document}

\maketitle

\label{firstpage}

\begin{abstract}
\normalsize
The effect of the expansion of powerful FR-II radio sources into a cluster
environment is discussed. The analysis considers both the thermal
and temporal evolution of the ICM which has passed
through the bow shock of the radio source and the effect of this swept-up 
gas on the dynamics of the radio source itself.
The final state of the swept-up ICM is critically dependent on the
thermal conductivity of the gas. If the gas behind the bow shock expands 
adiabatically, and the source is expanding into a steeply falling atmosphere,
then a narrow dense layer will form as
the radio source lifts gas out of the cluster potential.
This layer has a cooling time
very much less than that of the gas just ahead of the radio source.
This effect does not occur if the thermal 
conductivity of the gas is high, or if the cluster atmosphere is shallow. 
The swept-up gas also affects the dynamics of the radio source
especially as it slows towards sub-sonic expansion.
The preferential accumulation of the swept-up gas to the sides of
the cocoon leads to the aspect ratio of the source increasing.
Eventually the contact surface must become Rayleigh-Taylor unstable leading
both to inflow of the swept-up ICM into the cavity created by
the cocoon, but also substantial mixing of the cooler denser swept-up
gas with the ambient ICM thereby creating a multi-phase ICM.
The radio source is likely to have a marked effect on the cluster
on timescales long compared to the age of the source.

\end{abstract}


\section{Introduction}
\normalsize
Observations with both the ROSAT and CHANDRA satellites have demonstrated
clearly that the presence of a powerful FR-II
radio source very significantly affects 
the ICM/IGM in the vicinity of the source.  The simplest model for 
such a radio
source has an over-pressured cocoon of jet material expanding supersonically
into the external gas; expanding ahead of the cocoon a bow shock sweeps up
the surrounding gas which then forms a 
shell between the bow-shock and the contact surface separating shocked gas and
cocoon material.  
In this model the cocoon should therefore be devoid 
of ICM/IGM and the strong shock should heat the ICM/IGM.  Some observations,
but not all, support this general picture.  
Both ROSAT and CHANDRA data for Cygnus A
(Carilli, Perley \& Harris 1994; Smith et al. 2002)
show cavities in the X-ray emitting gas coincident with the radio cocoon and
the effects of the bow shock can be clearly seen.  
The observations of both the A426/3C84 interaction (Perseus; 
Fabian et al. 2000;
Fabian et al. 2002) and A2052/3C317 show clear evidence for
a radio cocoon devoid of X-ray emitting gas although in both cases
the radio structure is complex (Rizza et al. 2000; Blanton et al. 2001).  
In two cluster/radio-source interactions in A133 and A2626 studied by 
Rizza et al. (2000)
using ROSAT the X-ray emitting gas may well occupy the same volume as
the radio-emitting plasma.
The interaction of FR-I or complex radio sources with their environment
has often been interpreted in terms of the buoyancy effects of the low-density
radio-emitting plasma as in M87 (B\"{o}hringer et al. 1995; Br\"{u}gen and
Kaiser 2001), Hydra A (McNamara et al. 2000a; David et al. 2001; Nulsen et al.
2002) and Centaurus A (Saxton, Sutherland \& Bicknell 2002).  A recent review
of the X-ray observational data on radio-source/cluster interaction is
given in McNamara et al. (2002b).  This interaction between radio source and
the ICM has led a number of authors to consider whether radio sources
could solve the so-called ``cooling-flow problem'' (e.g. Binney and Tabor 1995;
McNamara et al. 2000b; Reynolds, Heinz \& Begelman 2001; Churazov et al. 2001;
Quilis, Bower \& Balogh 2001; B\"{o}hringer et al. 2002), in which the cool
gas expected to accumulate at the cluster centre as a result of cooling is not
detected.

A number of authors have considered various aspects of the physics and
evolution of the gas swept-up by the passage of the radio source.  
Scheuer (1974) pointed out the potential
for the swept-up gas to become Rayleigh Taylor unstable given that the
density of the cocoon must be very much less than that of the external
gas.  Gull \& Northover (1973) had already
considered the evolution of a light bubble of gas in a cluster potential
as a possible model for radio source confinement, while Smith et al. (1983) 
determined the conditions under which a bubble, rather than a jet, would be
produced in the Blandford and Rees (1974) twin-exhaust model.
This picture has
recently been updated by Churazov et al. (2001), 
Br\"{u}ggen \& Kaiser (2001), Quillis, Bower \&
Balogh (2001) and Saxton, Sutherland \& Bicknell (2002)
who have investigated the
evolution of buoyant bubbles.  These studies show that the
buoyant material can give rise to a mushroom-type structure
which although susceptible to Rayleigh-Taylor instability can significantly 
affect a cooling flow within the cluster.
Heinz, Reynolds \& Begelman (1998; hereafter HRB), calculated the
expected X-ray emission from the swept-up ICM using a simplified geometry
for the radio source and assuming uniform conditions between the
bow-shock and the contact surface.  A more detailed calculation 
of the heating effects of a radio source on its environment
was made by
Kaiser \& Alexander (1999; hereafter KA99) assuming an elliptical
source shape, self-similar evolution and adiabatic conditions
in the swept-up gas.  
They predict significant structure in the X-ray emission
behind the bow shock and flow in this region away from the
hotspot.  Reynolds, Heinz \& Begelman (2001) present detailed
simulations of a radio source expanding into an atmosphere;
for sources expanding supersonically their results are in agreement
with the analytical models of KA99, but they are able to follow
the evolution of the radio source into a phase in which it reaches pressure
balance with the surrounding gas and again find significant structure in the 
expected X-ray emission.  Despite this literature a number of important
problems concerning the evolution of the swept-up gas have not 
been fully addressed.
In particular what is the long term effect of the radio source 
on its environment
on time-scales long compared to the source life time, and what effect does the
swept-up gas have on the evolution of the radio source.  
Heating of the ICM/IGM by
radio sources may have important and measurable cosmological
effects (e.g. Yamada \& Fujita 2001).

In this paper I extend this earlier work to determine the effects of the
expansion of a powerful FR-II radio source into a cluster-like environment.
The structure of this paper is as follows.  In Section 2 I outline the main
physical considerations appropriate to the analysis of the evolution of
gas swept-up by the radio source.  In Section 3 I consider the evolution of
this gas in detail and in Section 4 I consider the effects of the gas on the
dynamics of the radio source.  Finally, in Section 5, I discuss the
implications of these results.

\section{General Considerations}
\normalsize
Before discussing more detailed models, it will be useful to consider the
main factors which are likely to be important for the swept-up gas and the
radio source itself.  I begin by reviewing the likely physical conditions
in the swept-up gas.  Figure 1 shows a sketch of the assumed geometry of the
source.
For simplicity I shall refer to the atmosphere into which the radio
source propagates as the Intra Cluster Medium (ICM).

As the radio source expands it must do work on the
surrounding medium.
For supersonic expansion the ICM will be heated
by the bow shock and is also lifted out of the potential well in which the
host galaxy sits; this swept-up gas resides between the bow shock and contact
surface.  
The sound speed in this shocked gas must be of order the velocity
of the bow shock itself, and therefore the sound crossing time must be somewhat
less than the age of the radio source suggesting that the pressure between
the bow shock and contact surface should be approximately constant.
This argument is confirmed by numerical simulations 
of FR-II radio sources (e.g. Reynolds et al. 2001)
and the calculations of KA99.  There must be pressure variations
within the shocked gas around the periphery of the source since the
pressure at the hotspot exceeds that within the cocoon; these pressure
variations drive a flow within the swept-up gas sending material away
from the hotspot in the direction of the host galaxy (KA99, Reynolds
et al. 2001).

As the ICM passes through the bow shock it is of course heated.
As this gas flows downstream of the bow shock it will evolve
adiabatically unless there is significant radiative heating /
cooling or the thermal conductivity is high.  The issue of the
cooling of the gas will be considered in detail below. 
The X-ray gas is optically thin, therefore it is only necessary
to consider the thermal conductivity
as a heating source. The thermal conductivity 
can be written in the form
$1.3 \eta n_{e}\lambda_{e}(kT_{e}/m_{e})^{1/2}$,
where the subscript $e$ refers to electron values, $\lambda_{e}$ is
the electron mean free path, and the dimensionless factor $\eta$ 
allows for suppression of the thermal conductivity 
below the canonical form given by Cowie \& McKee (1977).
The heating time, $t_{h}$, for a temperature variation of
$\Delta T$ to be removed in a layer of gas of thickness $\Delta r$ 
is given by $t_{h} \approx 1.5 n k \Delta T \Delta r / q$, 
where the heat flux, 
$q = 1.3 \eta n_{e}\lambda_{e}(kT_{e}/m_{e})^{1/2} \Delta (k T) / \Delta r$.
This simplifies since 
$(kT_{e}/m_{e})^{1/2} \sim \bar{v}_{e} \sim c_{s}(m_{p}/m_{e})^{1/2}$,
hence $t_{h} \sim 0.03 t_{s} (\Delta r / \lambda_{e}) / \eta$, where
$t_{s}$ is the sound crossing time across the region of size $\Delta r$.
The electron mean free path is given approximately by:
\[
\lambda_{e} \approx 23 
	\left(\frac{T_{e}}{10^8 \mathrm{K}}\right)^2 
	\left(\frac{n_{e}}{10^{-3} \mathrm{cm}^{-3}}\right)^{-1} \mathrm{kpc}.
\]
For Cygnus A at a radius of approximately 100~kpc 
$n_e \sim 7 \times 10^{-3}$cm$^{-3}$,
$T_e \sim T_{\mathrm{gas}} \sim 7 \times 10^7$K, 
$\Delta r \sim 10 \mathrm{kpc}$
(Smith et al. 2002), and therefore
\[
\frac{t_h}{t_s} \approx \frac{0.3}{\eta}.
\]
For the heating time of order the sound crossing time, 
and both much less than the age of the source, the swept-up layer of
gas should be both isothermal and isobaric. However, if the thermal
conductivity is significantly suppressed over the canonical form
given by Cowie \& McKee then the gas will evolve adiabatically
behind the shock.  Recently, Ettori \& Fabian (2000) have argued, based
on the analysis of the X-ray data for Abell 2142, that the thermal
conductivity is very significantly suppressed giving $\eta \sim 1/250$ to
as low as $\sim 1/2500$ at least in some regions of the cluster.
If the suppression of the thermal conductivity is of this order in all
cases then the gas behind the shock should clearly be treated adiabatically.
Given the remaining uncertainly in the value for the thermal
conductivity, 
I shall discuss both the adiabatic and isothermal cases below.
The temperature of the swept-up gas will be determined by dynamical
processes (e.g. shock heating, adiabatic expansion) and radiative cooling.
I will refer to the former processes as dynamical heating/cooling.

The mass of swept-up gas affects the dynamics of the radio source
in two ways: firstly the inertia of the gas must be considered and
secondly the gravitational force on the gas due to the cluster
potential.  As I will show below the inertia of the gas does not
alter the dynamics in the sense that the radio source may 
expand in a self-similar fashion as shown by Kaiser \& Alexander
(1997; hereafter KA97) -- their
model implicitly allows for the inertia of the gas.  A simple
calculation demonstrates that the gravitational force of the swept
up gas will also be important towards the end of the life time of the
source.  As the source expansion slows the cocoon pressure drops
towards pressure equilibrium with the confining atmosphere.  For
gas in a dark matter dominated cluster with no pre-heating of the
ICM, the pressure in the gas is simply determined by the requirement
for hydrostatic equilibrium in the potential well, hence the
pressure in the gas must be comparable to the gravitational 
potential energy
per unit volume of gas at any point in the cluster.  This
gravitational force must therefore be at least as important as the pressure
in the atmosphere as the source moves towards sub-sonic expansion.

Throughout this paper I will use the following conventions: subscripts
$j$, $c$, $h$, $s$ will refer to the jet, cocoon, hotspot and swept-up
gas respectively; the parameters in the external gas will be denoted by
the subscript $x$.
The external atmosphere is assumed to follow a King profile of the form
\[
\rho_x = \left(1 + (r/a_0)^2\right)^{-3\beta_K / 2}
\]
which is approximated as a piece-wise power-law (Alexander 2000),
$\rho_x = \rho_0 (r/a_0)^{-\beta}$,
with $\beta \sim 0$ for $r < a_0$ and $\beta = 3\beta_K$ for $r \gg a_0$.
Finally, I assume adiabatic indices for the ICM of $5/3$ and for the
radio-emitting plasma of the cocoon of $4/3$.

\section{Evolution of the swept-up gas}

\begin{figure}
\centerline{\epsfig{figure=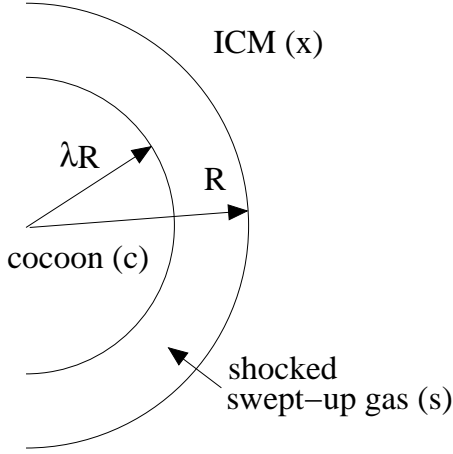,angle=0.0,width=6 truecm,clip=}}
\caption{The simplified geometry assumed for models of the
cocoon expansion}
\end{figure}

The discussion of Section 2 shows that it is the expansion of the
cocoon which is most important in determining the evolution of the
swept-up gas.  A simplified model is then sufficient to investigate this
evolution and derive scaling relationships and I therefore approximate the
expansion of the cocoon as a spherical expansion following the treatment
by HRB.

The modelled geometry of the cocoon and bow shock is shown in Figure 1.
The radius of the bow-shock is $R$ 
and that of the cocoon (or equivalently the contact surface) $\lambda R$.
The dynamics of the cocoon are determined by the energy input from the
jet at a constant rate $Q_0$.  
As shown in KA97 and by HRB
for this simplified geometry (see also Section 4.2)
the bow-shock and cocoon expand self-similarly with $\lambda$ a
constant and the radius of the bow shock given by 
$R = C a_0 \left(\frac{t}{t_0}\right)^{\frac{3}{5-\beta}}$, where
the characteristic time $t_0 = \left( a_{0}^{5}\rho_0 / Q\right)^{1/3}$.

Two related parameters which characterise the nature of the swept-up gas
are: (1) the ratio of the cooling time in this gas compared to that
of the ICM just ahead of the bow shock, and (2) the ratio of volume 
emissivities due to thermal Bremsstrahlung
for the swept-up gas and un-shocked ICM.  The former depends on the bolometric
luminosity and therefore, for gas of order a few $10^7$K, scales as 
$T^{1/2}/\rho$, whereas the appropriate emissivity for a relatively
narrow band detector is the specific emissivity and therefore scales
approximately as $\rho^2 T^{-1/2}$.

\subsection{The isothermal case}
I consider first the case when the gas between the bow-shock and contact
surface is taken to be isothermal; this assumption was used by 
HRB who then calculated the X-ray
emissivity for a spherical source expanding in a King profile.
Since the pressure is approximately
constant through the swept-up gas (Section 2) the density, $\rho_s$,
is uniform and
can be found simply from the total swept-up mass, 
$4 \pi R^{3-\beta} \rho_0 a_{0}^{\beta} / (3-\beta)$, and the volume
between bow shock and contact surface, $4 \pi (1-\lambda^3) R^3 / 3$.
The pressure in the swept-up gas follows from the jump conditions
at the shock which I shall assume to be a strong shock, 
$p_s = \frac{5}{4} M_{0}^{2} p_x = \frac{3}{4} \rho_x \dot{R}^2$, 
where a dot denotes differentiation
with respect to time, $\rho_x = \rho_0 (R/a_0)^{-\beta}$ is the
density just ahead of the shock at a radius $R$,
and $M_0$ is the Mach number of the bow-shock relative to the
ICM for a source of radius $R$.  
The temperature in the
swept-up gas, $T_s$, follows from the ideal gas law, $p = \rho k T / \mu$, where
$\mu$ is the mass per particle:
\[
T_s = \frac{\mu p_s}{k \rho_s} = \frac{5}{4}M_{0}^{2}\frac{\rho_x}{\rho_s} T_x.
\]

The cooling time of both the swept-up gas and the ICM just ahead of the
shock are proportional to $T^{1/2}/\rho$, hence the ratio of the cooling
time in the swept-up gas, $t_{cs}$ to that of the ICM just ahead of the shock,
$t_{cx}$ is
\[
\frac{t_{cs}}{t_{cx}} = \frac{T_{s}^{1/2}}{\rho_s}\frac{\rho_x}{T_{x}^{1/2}} =
\frac{\sqrt{5}}{2}M_0\left(\frac{\rho_x}{\rho_s}\right)^{3/2}.
\]
The density in the swept-up gas is given by 
$\rho_s = \frac{3}{(3-\beta)(1-\lambda^3)}\rho_x$, hence:
\[
\frac{t_{cs}}{t_{cx}} = \frac{\sqrt{5}}{2}\left((3-\beta)(1-\lambda^{3})\right)^{3/2} M_0.
\]
$(1-\lambda^3)$ is equal to $15/4(11-\beta)$ (Section 4.2),
hence $\frac{t_{cs}}{t_{cx}} > 0.05 M_0$ since $\beta < 2$.  
Except for high Mach numbers the cooling time in the swept-up gas is less 
than that
in the external atmosphere ahead of the shock, and this will certainly be
the case as the source slows to near sub-sonic expansion.

Similarly, the ratio of specific volume emissivities is given by:
\[
\frac{\epsilon_{vs}}{\epsilon_{vx}} = 
   \frac{T_s^{-1/2}\rho_s^2}{T_x^{-1/2}\rho_x^2} =
   \frac{2}{\sqrt{5}}\left(
	\frac{4}{5}\frac{3}{(3-\beta)(1-\lambda^3)}
   \right)^{5/2} M_0^{-1},
\]
and $\frac{\epsilon_{vs}}{\epsilon_{vx}} < 125 M_0^{-1}$.  
These results are illustrated in Figure 2.
\begin{figure}
\centerline{\epsfig{figure=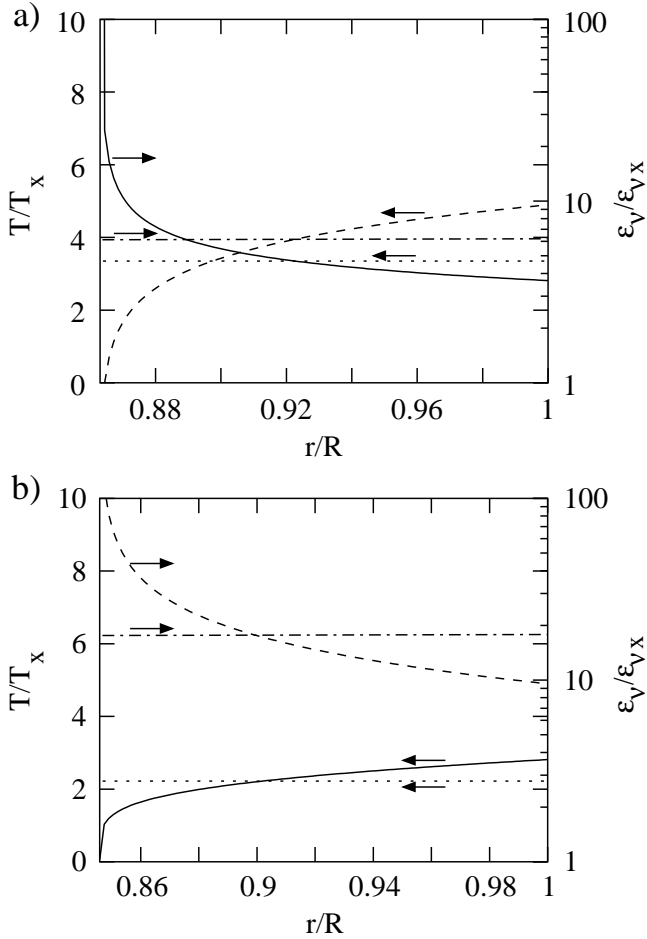,angle=0.0,width=8.5 truecm,clip=} }
\caption{The radial dependence of the temperature and volume emissivity 
in the region between the contact surface and the bow shock.  The two panels 
show the results for two different atmospheres (a) $\beta = 0.5$ and 
(b) $\beta = 1.5$.  In both (a) and (b) the lines have the following meaning:
solid line $T/T_x$ adiabatic; dotted line $T/T_x$ isothermal;
dashed line $\epsilon_{v}/\epsilon_{vx}$ adiabatic;
dashed-dot line $\epsilon_{v}/\epsilon_{vx}$ isothermal}
\end{figure}

The enhanced
emissivity of the swept-up gas over the ICM ahead of the shock leads
to an X-ray bright shocked shell as discussed by HRB.  
The mean isothermal temperature in the
shocked gas differs slightly from the post-shock temperature by a factor 
$5(3-\beta)/(11-\beta)$ in this approximation, however even in a steep
atmosphere the factor is close to unity, hence the swept-up gas never cools
due to dynamical cooling significantly below the temperature of the
gas ahead of the shock.  In the later stages of expansion
as the Mach number tends to unity, the cooling time
can become short compared to that in the cluster gas -- if the
timescale is comparable to the age of the source then significant
radiative cooling of the swept-up gas can occur.  Taking a source age
of $10^7$yrs and a steep atmosphere, then the cooling time in the
cluster gas would have to be less than of order a few $10^8$yrs for 
the swept-up gas to suffer significant radiative losses.  Such conditions
are only possible in the centre of the strongest cooling flow
clusters.  For sources still within the core radius of the cluster
the cooling time in the ICM would have to be nearly an order of
magnitude less.  It seems likely therefore that if the gas is
isothermal radiative cooling of the swept-up gas is negligible
over the lifetime of the source.

\subsection{The adiabatic case}
The situation is very different if the thermal conductivity is sufficiently
suppressed so that the swept-up gas evolves adiabatically.
Consider gas which passed through the shock when the bow-shock was of 
radius $R y$,
where $R$ is the radius of the bow shock at the current epoch and $0 \le y \le 1$.
The immediate post-shock conditions result in a density
and temperature ($\rho_y$, $T_y$) which are simply related by the
jump conditions to the density and temperature in the ICM; the density 
across the shock rises by a factor of 4, and the temperature increases 
by a factor of $\frac{5}{16}M_{y}^2$, 
where $M_y$ is the Mach number of the shock
at a radius $R y$.  As the source evolves to a size $R$ this gas must expand and
will cool adiabatically.
The density and temperature of this gas 
when the source
reaches a size $R$ ($\rho_s(y)$, $T_s(y)$) are given by the usual relationships
\[
\frac{\rho_s(y)}{\rho_y} = \left(\frac{p_s}{p_y} \right)^{1/\gamma}
\]
\[
\frac{T_s(y)}{T_y} = \left(\frac{p_s}{p_y} \right)^{1-1/\gamma}
\]
where $\frac{p_s}{p_y}$ is the ratio of the post shock pressure 
for cocoon radii of $R$ and $R y$.
In the self-similar model of KA97 (see HRB; and Section 4.2)
this pressure ratio has a power-law dependence on the source size of
\[
\frac{p_s}{p_y} = y^{\frac{4+\beta}{3}}.
\]
Assuming an isothermal atmosphere, the Mach number at radius $R y$ is related in
the self-similar model to the Mach number at radius $R$ by 
$M_y = M_0 y^{(\beta - 2)/3}$.
Using these results it follows that the density and temperature at the
present epoch for gas swept
up when the source was of size $R y$ are given in terms of the external
conditions at a radius $R$ as follows:
\[
\rho_s(y) = 4 \rho_x y^{\frac{4(1-\beta)}{5}}
\]
\[
T_s(y) = \frac{5}{16} T_x M_0^2 y^{\frac{4(\beta-1)}{5}}.
\]
The distribution of density and temperature between the contact surface and 
bow shock at the present epoch follow from conservation of mass.
The mass swept up between radii $R y$ to $R(y + dy)$ must now lie between
radii $R x$ to $R(x + dx)$ such that:
\[
4 \pi x^2 dx R^3 \rho_s(y) = 4 \pi y^2 dy R^3 \rho_x y^{-\beta}.
\]
Substituting the above results for the density gives a simple differential
equation relating $x$ and $y$ which has the solution
\[
x^3 = \frac{15 y^{(11-\beta)/5}}{4(11-\beta)} + \lambda^3
\]
which is valid for $\lambda \le x \le 1$ and $\lambda^3 = 1-15/4(11-\beta)$.
These solutions are illustrated in Figure 2.

These results are in agreement with the numerical calculations of HA99 and
the adiabatic wind models developed by Dyson, Falle \& Perry (1980).
It follows that the cooling time of this gas compared to the cooling
time of the gas immediately ahead of the shock at radius $R$ is given by
\[
\frac{t_{cs}(y)}{t_{cx}} = 
\frac{\sqrt{5}}{16} M_0 y^{\frac{6(\beta-1)}{5}}.
\]
For $\beta \le 1$ the cooling time of the post-shock gas will, for even very
modest Mach numbers, exceed that of the gas at a similar radius in the ICM.
For $\beta > 1$ some fraction of the gas will have a cooling time less than
the ICM at that same radius.  For example, taking $\beta = 1.5$, $M_0 = 10$,
all gas swept up when $y < 0.54$ will have a cooling time less than the
ICM at a radius $R$, or approximately 40\% of the swept-up gas by mass.  

The ratio of the specific volume emissivities follows similarly:
\[
\frac{\epsilon_{vs}(y)}{\epsilon_{vx}} = 
   \frac{T_s(y)^{-1/2}\rho_s(y)^2}{T_x^{-1/2}\rho_x^2} =
   \frac{64}{\sqrt{5}}y^{2(1-\beta)} M_0^{-1}.
\]
Unlike the isothermal case these results depend critically on the
form of the atmosphere into which the source is expanding and when the
gas was swept up.  Immediately behind the bow shock the X-ray emissivity
will be enhanced (unless the expansion is highly supersonic).  Between
the bow-shock and the contact surface the emissivity will fall for $\beta < 1$,
whereas for $\beta > 1$ the emissivity  rises
towards the contact surface, the temperature falls 
and the cooling time decreases.  For a steep 
atmosphere we therefore expect to find a layer of cooler, dense X-ray
luminous gas adjacent to the contact surface.

\section{Effects of the swept-up gas on radio source dynamics}

I have already discussed in Section 2 how the gravitational
force of the swept-up gas must become comparable to the cocoon
pressure as the expansion of the radio source slows to being
only mildly supersonic.  In this section
I consider in more detail the evolution of the radio source
itself.  There are three aspects that must be considered.
As first pointed out by Scheuer (1974), the contact
surface is liable to Rayleigh Taylor instability unless the
deceleration of the radio source is sufficient to stabilise
this interface.  I shall argue in Section 4.1 that where
observations suggest that the X-ray emitting gas is excluded from
the cocoon this stability requirement may place a useful additional
constraint on the dynamics of the radio source.
Even before the source approaches pressure equilibrium the
swept-up gas modifies the dynamics of the radio source
principally by leading to a departure from self-similar evolution
since the swept-up gas accumulates preferentially around the cocoon
and not the hotspot.  In Section 4.2 I develop a simple dynamical
model for this phase of the radio source evolution.  Finally in Section
4.3 I consider the later stages of evolution when the source expansion
becomes only mildly supersonic.

\subsection{Rayleigh-Taylor stability of the contact surface}

The contact surface between cocoon and the swept-up gas is liable
to become Rayleigh-Taylor (RT) unstable since the density of the swept
up material exceeds that within the cocoon.  The surface is
stabilised by the deceleration of the contact surface (Scheuer 1974).
If the total mass of gas and dark matter within a radius $r$ is $M(r)$,
and assuming spherical symmetry for the cluster, then the
component of the 
gravitational acceleration perpendicular to the contact surface
at a radius $r$
from the centre of the cluster is $g_{\bot} = GM(r)\cos(\psi)/r^2$, 
where $\psi$
is the angle between radius vector and the normal to the cocoon.
The deceleration of the contact surface cannot be 
measured directly from
observations, but can estimated in a (weakly) model dependent manner 
by assuming a scaling relation for the advance of the contact surface
of the form $R_c = \alpha t^{\delta}$, where
$R_c$ is an appropriate measure of the size of the cocoon.  
It follows that the
acceleration of the contact surface can be written:
\[
\ddot{R_c} = \delta^3 (\delta - 1) \frac{\dot{R_c}^2}{R_c}.
\]
For the self-similar model of KA97 the exponent $\delta = 3/(5-\beta)$,
and $R_c$ and $\dot{R_c}$, can both be estimated from observational
data; for the advance of the hotspot $R_c$ is simply the length of the
source, or for the cocoon it is the radius of the cocoon assuming
approximately cylindrical geometry.

Observations with CHANDRA show the cocoons of a number of radio sources,
in particular Cygnus A, to be devoid of X-ray emitting gas.  For the
contact surface to exclude the swept-up gas the surface must remain stable
to RT instabilities otherwise fast-growing short wavelength modes would
disrupt the contact surface allowing the swept-up mass to fill the cocoon
cavity.  The requirement for stability at any point on the contact surface
requires the surface to be decelerating and:
\[
| g_{\bot} | < | \ddot{R_c} | = | \delta^3 (\delta - 1) \frac{\dot{R_c}^2}{R_c} |.
\]
Smith et al. (2002) give data for the total mass (dark matter and gas)
within a radius $r$ in the Cygnus A cluster.  Using these results the
perpendicular component of the
gravitational acceleration at the hotspot of Cygnus A is approximately
$7.4 \times 10^{-10}$ms$^{-2}$, and for a point on the cocoon
approximately three-quarters along the lobe 
$1.4 \times 10^{-10}$ms$^{-2}$.  For self-similar models the
magnitude of the deceleration decreases as the atmosphere becomes
steeper.  To obtain an upper estimate for the deceleration I therefore
adopt a lower limit to $\beta$ of $\sim 1.5$ (Carilli et al. 1994;
Smith et al 2002), giving $\ddot{R_c} = 0.09 \dot{R_c}^2 / R_c$.
The hotspot advance speed can be estimated from spectral ageing arguments 
assuming an equipartition field strength within the lobes.
This gives for Cygnus A $\dot{L_j} \sim 0.05$c
(Alexander Brown \& Scott 1984) and since $L_j \sim 100$kpc
\[
|\ddot{L_j}| = 6.75 \times 10^{-9} \left(\frac{v}{0.05\mathrm{c}}\right)^2 
\left(\frac{L_j}{100\mathrm{kpc}}\right)^{-1} \mathrm{m}\mathrm{s}^{-2}.
\]
For self-similar expansion the acceleration of the side of the cocoon
is related to the hotspot acceleration via 
$\ddot{r_c} = \frac{r_c}{L_j}\ddot{L_j}$, giving for the cocoon
$\ddot{r_c} \sim 0.25 \ddot{L_j}$.  
The lobe speed derived from spectral ageing is almost certainly an upper limit.
Taking into account all of the dynamical constraints Alexander \& Pooley (1996)
estimate $\dot{L_j} \sim 0.005$c.  At best therefore the contact surface at the
hotspot of Cygnus A is likely to be just RT stable.  I will return to this
point in Section 5.

In principle, this method provides a way of obtaining a lower-limit to
the advance speed of a radio source when there is evidence that the
contact surface is still RT stable; the method relies only upon the existence
of good X-ray data and a simple, but robust, model for the source expansion.

\subsection{Dynamical model with swept-up gas}

In this section I investigate whether the swept-up mass has any effect
on the dynamics of the source before the source ceases to expand
supersonically.  KA99 show that there is significant flow in the swept
up gas driven by the pressure gradient between the hotspot and the
sides of the cocoon.  A simplified model for the dynamics is therefore
suggested in which there is no significant effect at the hotspot and
only the dynamics of the lobe are affected by the gravitational force
on the swept-up gas in the cluster potential.  
In order to make progress I therefore adopt
a simplified approach in which I treat the expansion of the cocoon as
purely spherical; this is the approach taken by HRB and
Reynolds \& Begelman (1997).  The expansion of the source in the hotspot
region will be similar to that discussed by Falle (1991) and KA97, and the
two parts of the overall solution can be related by requiring that
the jet is in pressure balance with the cocoon as in KA97.

The governing equations are similar to those given by HRB,
although I now allow for both the inertia and gravitational force
on the swept-up mass and use a more accurate treatment of the jump conditions
at the shock.  The dynamical analysis assumes the gas behaves adiabatically
throughout. 
As before, the radius of the bow shock is defined as $R$ and the radius of the
cocoon is written as $R_c = \lambda R$.
The first equation gives the conservation of
energy in the cocoon assuming that negligible energy is stored in the
hotspot:
\[
\frac{1}{\gamma_c - 1} V_c \frac{d p_c}{dt} + 
\frac{\gamma_c}{\gamma_c -1} p_c \frac{d V_c}{dt} = Q_0.
\]
Assuming a uniform pressure in the swept-up layer, energy conservation
in this gas gives
\[
\frac{1}{\gamma_s - 1} V_s \frac{d p_s}{dt} + 
\frac{\gamma_s}{\gamma_s -1} p_s \frac{d V_s}{dt} =
\frac{15}{16}\frac{4\pi R^2}{2} \rho_0 \left(\frac{R}{a_0} \right)^{-\beta} \dot{R}^3
\]
where the term on the right hand side is just the rate of heat input
at the bow shock (the factor of $15/16$, which differs from HRB,
allows for the bulk kinetic energy of the gas post-shock).
The approximation of a uniform pressure between bow shock and contact
surface is not
exact when the inertia of the swept-up gas is not negligible.  To simplify the
problem I assume, given the results of Section 3, that most of the
mass resides in a relatively thin layer either at the contact surface when $\beta > 1$,
or just behind the bow shock when $\beta < 1$.  Across 
the rest of the region between bow shock and contact surface I assume
a uniform pressure of $p_s$.  The pressure difference between cocoon and
the shocked gas can be equated to the acceleration of this narrow dense
layer plus the gravitational acceleration.  For $\beta > 1$
\[
4\pi \lambda^2 R^2  (p_c - p_s) = 
M_s \left( \ddot{(\lambda R)} + g_{D}(\lambda R) \right)
\]
Where $M_s = \frac{\rho_0 a_0^{\beta}}{3-\beta} R^{3-\beta}$ is the
swept-up mass, and $g_{D}(r)$ is the gravitational acceleration 
towards the centre of the cluster at radius 
$r$, due mainly to the dark matter in the cluster.
When $\beta < 1$ this equation is modified in the sense
that $\lambda R$ is everywhere replaced by $R$.
This equation differs from the equivalent form given by HRB
by inclusion of both an inertia term and the gravitational acceleration;
as outlined below the inclusion of the inertia term still permits a
self-similar solution to be found when the gravitational term is
negligible (KA97 implicitly allow for the inertia of the swept-up
gas in their dynamical solution).

The gravitational acceleration is found in a self-consistent manner by 
assuming that the distribution of dark matter is such as to maintain the
gas in hydrostatic equilibrium; for an isothermal gas distribution it 
follows that the gravitational acceleration is given by:
\[
g_D =  \frac{kT_x}{\mu} \frac{1}{\rho_x} \frac{d\rho_x}{dr},
\]
and for a King profile
\[
g_D =  3 \beta_K \frac{kT_x}{a_0\mu} \frac{r}{a_0} 
\frac{1}{\left[1 + \left(r/a_0\right)^2 \right]}
\]
which can be approximated as follows:
\[
g_D \approx X_D \left\{
  \begin{array}{ll}
	\left(\frac{r}{a_0}\right)^{-1} & r \gg a_0 \\
	\left(\frac{r}{a_0}\right) & r \ll a_0 \\
  \end{array}
\right.
\]
where $X_D = 3kT_x \beta_K/\mu a_0$.  

The final relationships needed
to close this set of equations are 
$V_c = 4\pi \lambda^{3}R^{3}/3$,
$V_s = 4\pi (1-\lambda^{3})R^{3}/3$,
and $p_s = \frac{3}{4}\rho_0\left(\frac{R}{a_0}\right)^{-\beta}\dot{R}^2$,
which follows from assuming the bow shock is a strong shock.  In this
expressions and throughout I will assume $\gamma_s = 5/3$ and $\gamma_c = 4/3$.

These equations can now be cast in dimensionless form by writing
$t = t_0 \tau$, $R = a_0 f(\tau)$, where $f$ and $\tau$ are the
dimensionless bow shock radius and time respectively.  Further, I shall
write $p_c = \zeta(\tau) p_s$ and note that given the constants of the
problem ($Q_0$, $a_0$ and $\rho_0$) the characteristic time
$t_0 = \left(\frac{a_0^5 \rho_0}{Q_0}\right)^{1/3}$. 
In terms of the dimensionless parameters the three governing equations
become:
\[
3\lambda^3 f^3 \frac{d}{d\tau}\left(\zeta \dot{f}^2 f^{-\beta} \right) +
4\zeta \dot{f}^2 f^{-\beta} \frac{d}{d\tau}\left(\lambda^3 f^3\right)
= \frac{1}{\pi}
\]
\begin{eqnarray*}
\lefteqn{
\frac{3}{2}(1-\lambda^3) f^3 \frac{d}{d\tau}\left(\dot{f}^2 f^{-\beta} \right) +
\frac{5}{2}\dot{f}^2 f^{-\beta} \frac{d}{d\tau}\left((1-\lambda^3) f^3\right)
=} \\
& & \frac{15}{8} f^{2-\beta} \dot{f}^3
\end{eqnarray*}
\[
\frac{3(3-\beta)}{4} \lambda^2 f^{2-\beta} \dot{f}^2 = 
f^{3-\beta} \left(
   \frac{d^2}{d\tau^2} (\lambda f)  + Y_D (\lambda f)^{\beta_D}
\right)
\]
where $Y_D = t_0^2 X_D / a_0$, $\beta_D = \pm 1$ and a dot indicates
differentiation with respect to $\tau$.  Again the third equation is written
for $\beta > 1$; for $\beta < 1$ there is no $\lambda$ dependence.
These equations admit a self-similar solution when the gravitational
term is negligible $Y_D \approx 0$ with $\lambda$ and $\zeta$ equal to
constant values ($\lambda_0$, $\zeta_0$)
and $f = C\tau^{\delta}$.  Substituting for this form
for $f$ gives, after a little algebra, $(1-\lambda_0^3) = \frac{15}{4(11-\beta)}$,
$\delta = \frac{3}{5-\beta}$, and 
$\zeta_0 = 1 - \frac{1-\delta}{\delta} \frac{4}{3(3-\beta)\lambda_0}$ for $\beta > 1$
and a similar expression omitting the $\lambda_0$ for $\beta < 1$.   The results
are in agreement with the results of KA97 and the previous section.

To investigate the case when $Y_D$ is small but non-zero, I seek a 
perturbed solution of the form $f = f_0(\tau)(1+F)$, 
$\lambda^3 = \lambda_0^3 (1+L)$ and $\zeta = \zeta_0(1+Z)$, where
$f_0(\tau) = C \tau^{3/(5-\beta)}$, the self-similar solution.
These trial solutions are then substituted into the original equations
and any terms which are quadratic or higher powers of $F$, $L$, $Z$,
$Y_D$ or any differential of one of these terms with respect
to $\tau$ are neglected.  The remaining equations are then linear
combinations of $F$, $L$, $Z$, $\tau\dot{F}$, $\tau^2\ddot{F}$, 
$\tau\dot{L}$ and $\tau^2\ddot{L}$, together with the gravitational
term which has the form $\tau^2 Y_D (\lambda_0 f_0)^{\beta_D-1}$.
A solution then exists such that each term ($F$, $L$, $Z$) has a 
power-law dependence on dimensionless time of the form $\tau^{\alpha}$,
and each term has the same time dependence as the perturbation,
$\tau^2 Y_D (\lambda_0 f_0)^{\beta_D-1}$, giving
$\alpha = 2 + 3(\beta_D - 1)/(5-\beta)$.
Substituting this form for the perturbation into the
original equations results in a set of 3 coupled linear algebraic
equations which can be solved to give the magnitudes ($F_1$, $L_1$,
$Z_1$) of each term (e.g. $F = F_1 \tau^{\alpha}$) 
in terms of the perturbation $Y_D$.
Further details are given in the appendix.
For large cluster radii, $\beta_D = -1$ and if the atmosphere is
steep $\beta \approx 2$, $\alpha \approx 0$ and we recover
a self-similar solution since the perturbation does not grow
with time.  

These solutions can be used to investigate how the aspect ratio
of the source, which I define here to be given by the ratio of the
jet to the radius of the cocoon, $L_j/R_c$, changes with time.
The time dependence of $R_c$ in terms of the perturbed solution
is straightforward $R_c/a_0 = \lambda f = \lambda_0 f_0 (1 + F + L/3)$
to first order.  To determine the time dependence of $L_j$ in
this approximation we return to the method employed by KA97
where the jet as it propagates through the cocoon is assumed to
be confined by the cocoon pressure, $p_j = p_c$.  KA97 show that
this implies that the hotspot pressure and cocoon pressure must have
the same time dependence; this result still holds in
the current approximation.  

Assuming no accumulation of material
at the hotspot, the advance of the hotspot is simply determined
by balancing the ram pressure to the hotspot pressure:
\[
\rho_0 \left(\frac{L_j}{a_0}\right)^{-\beta} \dot{L_j}^2 = p_h \propto p_c
\]
where the time dependence of $p_c$ must be given by the perturbed
solution.  For an approximately power-law dependence of $L_j$ on time,
$\dot{L_j} \sim L_j / (t_0 \tau)$.  The cocoon pressure depends on
$\zeta f^{-\beta} \dot{f}^2$, hence writing the time dependence
to first order
\[
L_j^{(2-\beta)\delta} \propto \tau^{(2-\beta)\delta -2}
\left(
   1 + (2-\beta)F + \frac{2\alpha}{\delta}F + Z
\right)
\]
and $R_c \propto \tau^{\delta}(1+F+L/3)$.
In Figure 3 I show how $L_j$ and $R_c$ evolve with time under the perturbation.
The cocoon radius changes slowly compared to the length of the jet; the cocoon
pressure in the perturbed solution exceeds that in the self-similar case at the
same time since there must now be a greater pressure difference between shocked gas 
and the cocoon to overcome the gravitational force on the swept-up gas and the 
cocoon volume is correspondingly smaller.  This increased lobe pressure acts to 
collimate the jet since the jet is assumed to be in pressure balance with the
cocoon material and the jet therefore propagates more quickly through the 
external atmosphere.
\begin{figure}
\centerline{\epsfig{figure=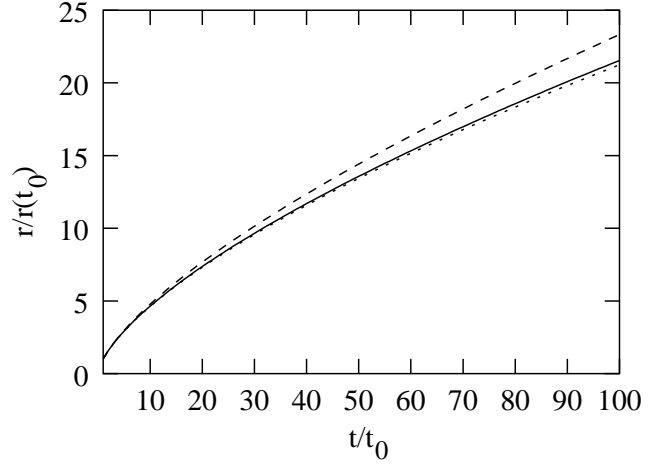,angle=0.0,width=8.5 truecm,clip=}}
\caption{The evolution of the jet length and cocoon radius for the
perturbed solution discussed in the text compared to the unperturbed solution.
The curves are normalized to the unperturbed solution at $t = t_0$.}
\end{figure}
The evolution of the
aspect ratio of the source from its unperturbed (similarity) value
follows from the above considerations and is given by:
\[
\frac{L_j}{R_c} = \left(\frac{L_j}{R_c}\right)_0
\left(
   1 - \frac{L}{3} + \frac{2\alpha F}{\delta(2-\beta)} + \frac{Z}{2-\beta}
\right).
\]
The perturbation term is proportional to
$Y_D \tau^{\alpha}$, and the associated coefficient is
found by solving the algebraic equations as given in the appendix.  
I now
consider these results for a typical case when $\beta_K = 0.5$,
and in the two limits when $L_j \ll a_0$ i.e. $\beta =0$ and $\beta_D = 1$,
and $L_j \gg a_0$ i.e. $\beta =3/2$ and $\beta_D = -1$:
\[
\frac{L_j}{R_c} \approx \left(\frac{L_j}{R_c}\right)_0 \times
\left\{
  \begin{array}{ll}
	1 + 0.2 Y_D \tau^2 & L_j \ll a_0 \\
	1 + 2.9 Y_D \tau^{2/7} & L_j \gg a_0 \\
  \end{array}
\right.
\]
These results can be cast in a more useful form as follows.  
The perturbation term has the form 
$Y_D = \frac{3}{4}\frac{kT_x t_0^2}{\mu a_0^2}$, which is
equal to $\frac{9}{20}\frac{c_{sx}^2 t_0^2}{a_0^2}$, where $c_{sx}$
is the (constant) sound speed in the ICM.
Furthermore, since 
$\dot{L_j} = \frac{3}{5-\beta}a_0 C t^{\frac{\beta-2}{5-\beta}} 
t_0^{\frac{-3}{5-\beta}}$,
$a_0$ can be re-written in terms of $\dot{L_j}$ to give:
\[
\frac{L_j}{R_c} \approx \left(\frac{L_j}{R_c}\right)_0 \times
\left\{
  \begin{array}{ll}
	1 + \frac{0.04}{M^2}\left(\frac{t}{t_0}\right)^{6/5} & L_j \ll a_0\\
	1 + \frac{0.9}{M^2} & L_j \gg a_0\\
  \end{array}
\right.
\]
In this expression $M$ is the Mach number of the hotspot advance
at the time $t$.  Although the perturbation grows faster for $L_j < a_0$,
the effect is never significant since the source reaches the
core radius at a time $t \sim t_0$.  
For $L_j \gg a_0$ the perturbation can be significant
especially as the source slows to being only mildly 
supersonic.  The perturbation grows most slowly for steep atmospheres
(as expected), however for the example considered here even for 
$\beta_K = 0.5$ the gravitational perturbation can significantly change
the aspect ratio of the source in the sense that older sources will
tend to be longer and thinner.  Of course, the considerations of the
previous section show that at some point the contact surface must become
RT unstable, and hence it is unlikely that in the majority of cases 
very long thin sources can be formed by this mechanism alone.

\subsection{The final stages of the cocoon}

The latter stages of development of the cocoon are the most
difficult to treat with simple analytical arguments.  The gravitational
force due to the swept-up gas becomes comparable
to the ram pressure as the source slows to being mildly supersonic.  

Of particular interest is the ultimate fate of the swept-up gas.
One possibility is that as the source slows the onset of the RT
instability (Section 4.1)
leads to this gas falling back through the contact surface
and refilling the low-density region previously occupied by
the cocoon.
Br\"{u}gen \& Kaiser (2001) have performed simulations 
to model the latter stages of evolution of a radio source by
considering of a light
bubble of plasma initially in pressure balance with the gas in the cluster.
They find very different behaviour depending on the initial shape of the
cloud.  Near spherical clouds are quickly destroyed by the RT instability
leading to mixing of the cocoon and ICM.  Elongated clouds can form a
mushroom-type feature which rises through the potential of the cluster
in a manner first suggested by Gull and Northover (1973).  In the wake behind
these mushrooms colder material is lifted out of the cluster.  In their
simulations  Br\"{u}gen \& Kaiser do not include the swept-up gas layer
I have discussed in this paper.  For the isothermal case of Section 3
when the source reaches pressure balance
this layer contains mostly gas with a shorter cooling time than the
ambient gas in the cluster.  The buoyant phase studied by Br\"{u}gen \& Kaiser
will then be very efficient at mixing this gas into the ambient material 
at large cluster radii.

Further discussion of the later stages of the source are beyond the scope
of the present paper.  We are currently undertaking simulations to
investigate further this phase of evolution.

\section{Discussion}

Given the results of the previous two sections it is now possible to
compare the model predictions with recent X-ray data and to
consider the evolution of the radio source and the cluster itself.

\subsection{Comparison to X-ray observations}

As discussed in the introduction there exist a number of studies of
the interaction of radio sources with their environment probed by
X-ray data.  In this section I will focus on those published studies
of interactions 
which have sufficient resolution and sensitivity to probe directly
the gas between bow-shock and contact surface.

The CHANDRA results for the Perseus cluster (Fabian et al. 2000;
Fabian et al. 2002) show clear evidence of a radio source
displacing the cluster medium.  Holes are observed in the
X-ray emission spatially coincident with the bright radio emission
from 3C84.  Surrounding these holes are bright X-ray rims which are
cooler than the surrounding gas.  Fabian et al. (2002) conclude that
the source must be expanding subsonically since the heating effect of
a strong shock is not observed.  The results of Section 3 provide an
alternative explanation for these data although the explanation
proposed by Fabian et al. cannot be ruled out.  For supersonic
expansion into a declining atmosphere under adiabatic
conditions I have shown that a layer of cool bright X-ray gas surrounds
the cocoon, although immediately behind the shock the gas temperature
is still increased.  The observational data will
effectively measure an emission weighted temperature and therefore
the cool bright rims surrounding the holes
can be interpreted as emission from a layer of cool gas adjacent 
to the contact surface.  A similar conclusion applies to the data for
Abell 2052/3C317 (Blanton et al. 2001)
in which cool X-ray rims are again observed surrounding holes
in the X-ray emission spatially coincident with the radio structure.  In this
case optical emission is also observed from the X-ray rims (Baum et al. 1988)
consistent with a cooling time in the swept-up gas which is significantly
reduced compared to that in the undisturbed ICM again consistent with the
analysis of Section 3.   Both of these comparisons should be
treated with some caution since the radio structure of both 3C84 and 3C317
is complex and does not resemble a classical FR-II, however 
the analysis presented here is still
applicable if the radio structures are surrounded by a bow shock.

The best test case is that of Cygnus A.
In this case the recent CHANDRA results presented by Smith et al. (2002)
permit an estimate of the mass of cool gas which may be present.  For a
hotspot advance speed 
such as that derived from spectral ageing 
of order $0.04$-$0.05$c (Alexander et al. 1984), 
and a ratio of cocoon radius to length of approximately 0.2 - 0.25,
the expansion velocity of the cocoon is of order $0.01$c.  Using the
data from Smith et al. at a radius typical of a point mid-way along the
cocoon ($\sim 70$kpc) the sound speed in the ICM is of order $0.003$c,
giving a Mach number for the sideways expansion of the cocoon of
about 3. The results of Section 3 suggest that all
gas swept up for $y \sim 0.2$ will now be cooler than the
gas just ahead of the bow-shock at a cluster radius of $\sim 70$kpc. 
Estimating the mass of gas this corresponds to is very uncertain.
The linear extent of Cygnus A in the cluster is of order $100$kpc and
the total gas mass within this radius is or order $10^{12}$M$_{\odot}$;
within this radius Cygnus occupies about 6\% of the volume giving
a total mass of swept up gas of order $6 \times 10^{10}$M$_{\odot}$.
Again, assuming a relatively steep atmosphere this suggests that 
approximately $10^{10}$M$_{\odot}$ of gas swept up by the radio source 
may have cooled to below the temperature of the ICM.  As discussed
by KA99, flow between the bow-shock and contact surface will lead to
accumulation of gas around the edge of the cocoon nearest to the
host galaxy and it is here that the cool gas should therefore be found.
The detailed calculation of KA99 when applied to Cygnus A also predicted
an annulus of cold gas in this location.  Interestingly, the
observations of Smith et al. (2002) show conclusive evidence for arcs
of cooler gas arranged in a belts around the host galaxy which 
are naturally explained as resulting from the accumulation of
adiabatically cooled swept-up ICM.
If the advance speed of the hotspot and cocoon are
significantly less than this value as suggested by considering the 
ram pressure within the hotspot and cocoon (Alexander \& Pooley 1996)
then the discussion of Section 4.1 shows that the contact surface 
may be RT unstable. If this is the case then the swept up gas may well
be being entrained within the cocoon.  This is most likely to occur in
the first instance to the sides of the cocoon where most cold material
has accumulated.  In this case the belts of cooler X-ray gas will be
falling through the contact surface and may, as suggested by by
Smith et al. (2002) provide a mechanism for continued fuelling 
of the AGN.

\subsection{Evolution of a radio source}

The radio source sweeps up a very significant mass of gas
during its lifetime which is displaced from the centre of the
cluster.  Depending upon the effectiveness of thermal
conduction this gas may cool adiabatically to form a thin dense
layer of relatively short cooling time (relative to the ambient
ICM at a similar distance from the cluster centre).  

Both the inertia and gravitational forces exerted by the swept-up
gas will modify the dynamics of the radio source.  Using the 
results above it is possible to determine in outline the likely
evolution of a powerful source.  The initial stage of evolution
will be as the source expands through the core of the cluster or
the ISM of the host galaxy.  In either case the mean density 
of the atmosphere will
be approximately constant.  During this phase the source will follow
a self-similar form of evolution as discussed by KA97 and Alexander (2000).
Although the effects of the swept-up mass are growing, they do
not become significant while the source is in this phase (Section 4.2).
As shown by KA99 there will be flow of the swept-up gas away from the
region ahead of the hotspot to the side of the source.
Despite this preferential accumulation of material at the 
surface of the cocoon self-similar expansion will still occur
since there is a feedback between the cocoon pressure and hotspot
pressure brought about by the jet reaching pressure balance with the
cocoon.

If the source is sufficiently powerful, and the jet is not
disrupted by a Kelvin-Helmhotz instability (Alexander 2000), the
source will reach a radius equal to the core radius of the 
atmosphere.  The hotspot clearly reaches the region of declining
density first, and as the source passes from the cluster core to
a declining atmosphere the aspect ratio is likely to increase
in a phase of non self-similar expansion.
The cocoon pressure will remain high determined by the properties
of the atmosphere in the cluster core, the jet pressure will therefore
also remain high and hence the hotspot pressure will exceed that 
predicted for self-similar expansion into a declining atmosphere.  
This phase will cease when the bow-shock fully
enters the declining atmosphere, and the source re-enters a phase
of self-similar expansion.  Beyond this point the effect of the
swept-up gas (Section 4.2) becomes important and as the Mach number
of the jet decreases the aspect ratio of the source will again increase.

Throughout this expansion the contact surface may be liable to RT instability.
This is most severe when the deceleration of the contact surface is 
small.  During the transition of the source through the cluster
core this is likely to be the case.  Any instability will allow 
swept-up gas to be entrained into the cocoon material and since the density
of this gas increases towards the centre of the source, it seems
likely that any instability will be most apparent in this region.
This process is also seen in the simulations of Reynolds et al. (2001).
The source may well survive this entrainment if the jet is not
disrupted, but this does form a natural mechanism for excluding the
cocoon from the region around the host galaxy.  

The final stages of the evolution of the source will be dominated by the
onset of the RT instability as the source slows.  The gravitational forces
on the swept-up gas will be largest near the host galaxy (where most of the
swept-up gas has accumulated).  These forces will lead to a pinching off
of the cocoon near the host galaxy and the onset of a buoyant phase
(Br\"{u}gen \& Kaiser 2001).
It may well be that the ingress of material at this time disrupts
the jet and effectively kills the source.  The buoyant phase will
result in the mixing of the remaining swept-up gas as discussed above
or, when the geometry is optimal, the formation of a true mushroom cloud
of radio-emitting plasma as discussed by Br\"{u}gen \& Kaiser (2001).
An alternative scenario is that the RT instability is effective
over the entire contact surface; in this case mixing of the radio emitting
plasma and swept-up gas will be very efficient and the cocoon will
be filled with X-ray emitting material with the radio-emitting
plasma having a non-unity filling factor.  
Pockets of low-density cocoon material will remain buoyant
leading to very efficient mixing.
 
The analysis of the previous two sections has considered a simple
model for the cluster atmosphere and in particular the relationship
between the swept-up mass and the depth of the potential well.  
If the cluster gas were pre-heated the gas pressure in
the atmosphere will dominate over gravitational forces of the
swept-up layer and the mass of this layer may be negligible.
In principle it may be possible to use the dynamics of radio sources
to provide an additional probe of the cluster environment,
however this requires an independent method of measuring the
pressure within the cocoon and expansion speed of the source.
The latter may be accessible to VLBI in the future, however at
present the equation of state of the radio-emitting plasma is very
uncertain (e.g. Hardcastel \& Worrall 2000) and this must be
resolved before radio sources can be used as 
detailed probes of their environment.

\subsection{The effect on the ICM}

It is now possible to consider the effect of the radio source on the ICM.
Whether the thermal conductivity is suppressed or not, the radio source must
do significant work on the ICM; the total $pdV$ work must be comparable
to the stored energy within the lobes (KA97, KA99), and this is deposited as
heat within the gas between the bow-shock and contact surface.  However,
the effect of the radio source is also to move cluster gas out of the centre 
of the cluster to large radii.  

In the early stages of evolution of the radio source we expect high Mach-number
expansion into a
relatively flat atmosphere.
Whether the thermal conductivity is suppressed or not the results derived above
show that the
swept-up gas is efficiently heated, a bright X-ray shell develops
between the cocoon and bow-shock and this gas has a cooling time
longer than the surrounding material.  
Furthermore, the radio source is
lifting a large mass of gas out of the centre of the cluster.  
Clearly during this phase the radio source halts any cooling flow in the cluster.

As the source expands into a declining atmosphere this picture
changes.
If the gas behaves adiabatically and the cluster atmosphere is relatively 
steep, $\beta > 1$, then the effect of the radio source is
to deposit a substantial amount of gas at large cluster radii which has a
cooling time less than the existing ICM at a similar radius.
Just ahead
of the contact surface there will a layer of cooler gas. In the model 
developed above there would apparently always be some gas at arbitrarily low
temperature, however this is clearly an artefact of assuming a power-law
atmosphere extending to the centre of the cluster.  In reality, this layer
will be thin, while the source is
expanding highly supersonically, however as the Mach number of the
expanding cocoon falls the mass of cooler gas will become significant.  
Even for what we think of as powerful supersonically expanding sources
a significant mass of gas may be in a cooler phase.  

As the source slows towards sub-sonic expansion the onset of RT
instabilities as discussed above will lead to a mixing of the
swept-up gas with the ambient gas and the cocoon material.
The effect of this process on the cluster may be dramatic.
If the swept-up gas has evolved adiabatically and the expansion
has been into a steep atmosphere then the effect is to
deposit gas at large cluster radii with a cooling time less
than the ambient material.  Indeed some gas adjacent to the
contact surface may be quite cool and have a very short cooling
time.  It seems likely therefore that the mixing process
driven by the buoyancy of the radio plasma will lead to a 
multi-phase ICM.
As the mixing proceeds this multi-phase gas must refill the
cavity occupied by the radio source.  
This will either be dominated by fluid-dynamical processes
or dynamical infall of the densest gas.  In either case
the timescale will be of order the sound crossing time, $t_s$,
across the radius of the cluster out to which the source expanded.
The age of the radio source however is of order $t_s / M$ where
$M$ is an average Mach number for the expansion.  Therefore the
timescale for the cluster gas to refill the cocoon must be of
order $M$ times the age of the source when it dies; this
can clearly be as long as an order of magnitude.  During this
phase their will be no distinct radio source (a cluster halo
source may possibly exist), but there will be a very significant
inflow of multi-phase cluster gas.  At the end of this
period the cluster within at least the radius reached by the
source will be filled with a multi-phase medium with a
cooling time in the outer parts of this region reduced over that
prior to the existence of the radio source.  

Similar conclusions concerning the effect of AGN heating were 
reached by Binney \& Tabor (1995); their analysis also
predicts suppression of the cooling flow followed by
a period of catastrophic cooling after the period of
AGN heating.  
The long term effect of a radio source
event on the cluster is however likely to be determined
by the buoyant rise of material through the cluster
atmosphere (Churazov et al. 2001; 
Br\"{u}ggen \& Kaiser 2001; 
Quillis, Bower \& Balogh 2001).  The subsequent evolution of 
this (no doubt turbulent)
multi-phase medium is difficult to determine without the
aid of simulations.  Whether a subsequent cooling flow is enhanced or
suppressed will be determined not only by the details of this
multi-phase ICM, but also by the epoch at which the cluster is
observed relative to the last radio-source event.

\section{CONCLUSIONS}

The effect of a radio source on the cluster gas and vice-versa
has been considered.  I have shown that, if the thermal
conductivity in the gas is suppressed as has been suggested
by recent observations with CHANDRA, then a cool dense layer 
of swept-up ICM material will form around the contact surface of
a powerful FR-II radio source expanding into a steeply
falling atmosphere.  This results simply from the
adiabatic expansion of dense gas as it is carried
out of the potential well of the cluster ahead of the
advancing cocoon of the radio source and only occurs
when the density of the cluster atmosphere falls
faster than $1/r$ for large cluster radii.
As the source slows from highly supersonic expansion to
near sub-sonic expansion almost all of the swept-up gas has
a cooling time less than the ambient ICM; some gas is likely
to be very cool and there is the possibility of promoting
the formation of a multi-phase medium.

Eventually the layer of swept-up gas must become Rayleigh-Taylor unstable.
The swept-up gas accumulates preferentially at the edges of the cocoon
away from the hotspot due to flow in the swept-up layer driven
by the pressure gradient which must exist due to the difference
in pressure between hotspot and cocoon.  Once the instability sets in
gas will fall through the contact surface and begin to fill the cavity
formed by the cocoon.  Using existing simulations as a clue to the
behaviour of the source in this phase it would seem that depending
on the detailed geometry of the source at this time there will
be both infall of material and mixing of the swept-up gas with the
ambient ICM.  

The long-term effect of the radio source on the cluster when there is
(a) suppression of the thermal conductivity and (b) efficient mixing
of the swept-up layer with the ambient gas, may well be to
reduce the cooling time of the gas at large cluster radii.  This
may promote the formation of a multi-phase ICM and/or for a period
of order a few times the lifetime of the source enhance or
instigate a cooling flow.  Certainly there must be a substantial
flow back into the cluster core as the swept-up mass falls through the
contact surface and fills the cocoon.

The swept-up ICM must also have a significant effect on the dynamics
of the radio source especially as the source slows
towards sub-sonic expansion; if there is no pre-heating of the cluster
gas then the gravitational force of the swept-up gas per
unit area of the cocoon must be at least comparable to the
external pressure.  For those sources which escape the core radius of
the cluster the gravitational force due to the swept-up gas 
can lead to an increase in the aspect ratio of the source as the
Mach number of expansion decreases.  For very steep atmospheres
falling nearly as fast as $1/r^2$ this perturbation does not
grow and the source can expand throughout its life in a self-similar
fashion.

\section*{Acknowledgments}
 
I thank Julia Riley, Malcolm Longair, Garrett Cotter 
and Jacques Basson for helpful discussions.  I also thank the referee 
for suggestions which substantially improved the paper.

\appendix
\section{The dynamical model}

In this appendix I give more details of the solution to the
dynamical equations governing the evolution of the cocoon.
The three governing equations in dimensionless form
are (see Section 4.2):
\begin{equation}
3\lambda^3 f^3 \frac{d}{d\tau}\left(\zeta \dot{f}^2 f^{-\beta} \right) +
4\zeta \dot{f}^2 f^{-\beta} \frac{d}{d\tau}\left(\lambda^3 f^3\right)
= \frac{1}{\pi}
\end{equation}
\begin{eqnarray}
\lefteqn{
\frac{3}{2}(1-\lambda^3) f^3 \frac{d}{d\tau}\left(\dot{f}^2 f^{-\beta} \right) +
\frac{5}{2}\dot{f}^2 f^{-\beta} \frac{d}{d\tau}\left((1-\lambda^3) f^3\right)
=}\nonumber\\
& &  \frac{15}{8} f^{2-\beta} \dot{f}^3
\end{eqnarray}
\begin{equation}
\frac{3(3-\beta)}{4} \lambda^2 f^{2-\beta} \dot{f}^2 = 
f^{3-\beta} \left(
   \frac{d^2}{d\tau^2} (\lambda f)  + Y_D (\lambda f)^{\beta_D}
\right)
\end{equation}
where $Y_D = t_0^2 X_D / a_0$ $\beta_D = \pm 1$ and a dot indicates
differentiation with respect to $\tau$.  The third equation as
written assumes $\beta > 1$; for $\beta < 1$ the equivalent equation
is found by replacing each occurrence of $\lambda f$ by $f$.

These equations admit a self-similar solution when the gravitational
term is negligible $Y_D \approx 0$ with $\lambda$ and $\zeta$ equal to
constant values ($\lambda_0$, $\zeta_0$)
and $f = C\tau^{\delta}$.  Substituting for this form
for $f$ in A1 gives
\[
\frac{1}{\pi C^{(5-\beta)}\lambda_0^3\zeta_0} = \frac{27}{(5-\beta)^3}(8-\beta)
\]
and from A2:
\[
1-\lambda_0^3 = \frac{15}{4(11-\beta)}
\]
and from A3:
\[
\zeta_0 = 1 - \left(\frac{1-\delta}{\delta}\right) \frac{4}{3(3-\beta)\lambda_0}
\]
and $\delta = 3/(5-\beta)$.  This result again holds for $\beta > 1$, for $\beta < 1$:
\[
\zeta_0 = 1 - \left(\frac{1-\delta}{\delta}\right) \frac{4}{3(3-\beta)}
\]

When $Y_D$ is small but non-zero, I seek a 
perturbed solution of the form $f = f_0(\tau)(1+F)$, 
$\lambda^3 = \lambda_0^3 (1+L)$ and $\zeta = \zeta_0(1+Z)$, where
$f_0(\tau) = C \tau^{3/(5-\beta)}$.  Substituting these trial solutions
into A1 to A3 and neglecting 
any terms which are quadratic or higher powers of $F$, $L$, $Z$,
$\dot{F}$, $\ddot{F}$, $\dot{L}$, $\ddot{L}$, 
or $Y_D$ gives:
\begin{eqnarray}
0 & = & 3[(2-\beta)\delta - 2]\left(L+Z+(5-\beta)F + \frac{2\tau}{\delta}\dot{F}\right) + \nonumber\\
& & 3\tau \left(\dot{Z} + (2-\beta)\dot{F}  + \frac{2}{\delta}\dot{F} + \frac{2\tau}{\delta}\ddot{F}\right) +\nonumber\\
& & 12\delta \left(Z+L+(5-\beta)F+\frac{2\tau}{\delta}\dot{F} \right) +\nonumber\\
& & 4\tau\left(\dot{L} + 3\dot{F}\right)
\end{eqnarray}
\begin{eqnarray}
\lefteqn{\frac{15}{8}\delta\left((5-\beta)F + \frac{3\tau}{\delta}\dot{F}\right) =} \nonumber\\
& & \frac{3}{2}\lambda_3[(2-\beta)\delta + 2]\left((5-\beta)F - l + \frac{2\tau}{\delta}\dot{F}\right) +\nonumber\\
& & \frac{3\tau}{2}\lambda_3\left((2-\beta)\dot{F} + \frac{2}{\delta}\dot{F} + \frac{2\tau}{\delta}\ddot{F}\right) +\nonumber\\
& & \frac{15\delta}{2}\lambda_3\left((5-\beta)F - l + \frac{2\tau}{\delta}\dot{F}\right) + \nonumber\\
& & \frac{5\tau}{2}\lambda_3\left(3\dot{F} - \dot{l}\right)
\end{eqnarray}
where $\lambda_3 = 1-\lambda_0^3$ and $l = L/\lambda_3$.
Equation A3 becomes
\begin{eqnarray}
\lefteqn{\delta(\delta-1)\left(\frac{Z}{\zeta_0-1}+\frac{2\tau}{\delta}\dot{F} - \frac{L}{3}\right) =}\nonumber\\
& & 2\tau\delta\left(\dot{F}+\frac{\dot{L}}{3}\right) + \tau^2\left(\ddot{F}+\frac{\ddot{L}}{3}\right) +
Y_D(\lambda_0f_0)^{\beta_D - 1}
\end{eqnarray}
for $\beta > 1$, and where $f_0 = C a_0 \tau^{3/(5-\beta)}$.
For $\beta < 1$ the terms involving $L$, $\dot{L}$ and $\ddot{L}$ do not appear
and $\lambda_0 f_0$ becomes simply $f_0$.
A solution then exists such that each term ($F$, $L$, $Z$) has a 
power-law dependence on dimensionless time of the form $\tau^{\alpha}$,
and this time dependence is the same as the perturbation,
$\tau^2 Y_D (\lambda_0 f_0)^{\beta_D-1}$, giving
$\alpha = 2 + 3(\beta_D - 1)/(5-\beta)$.

Substituting this form of the solution with
$F = F_1 \tau^{\alpha}$ etc. we obtain a set of
algebraic equations for the
amplitudes $F_1$, $L_1$ and$K_1$.  
\begin{eqnarray}
\lefteqn{
\left[(18-3\beta)\delta - 6 + 3\alpha\right]\left(Z_1+L_1+(5-\beta)F_1 + \frac{2\alpha}{\delta}F_1\right)
} \nonumber\\
& & {} + \alpha\left(L_1 + 3F_1\right) = 0
\end{eqnarray}
\begin{eqnarray}
\lefteqn{
\frac{15\delta}{4\lambda_3}\left[(5-\beta) + \frac{3\alpha}{\delta}\right] F_1 =
2\alpha\left(3F_1-l_1\right) + }\nonumber\\
& & 3[\alpha+(7-\beta)\delta-2]
\left((5-\beta)F_1-l_1+\frac{2\alpha}{\delta}F_1\right)
\end{eqnarray}
\begin{eqnarray}
\lefteqn{
\delta(\delta-1)\left(\frac{Z_1}{\zeta_0-1}+\frac{2\alpha F_1}{\delta}-\frac{L_1}{3}\right) =}\nonumber\\
& & [2\alpha\delta + \alpha(\alpha-1)]\left(F_1 + \frac{L_1}{3}\right) +
Y_D\lambda_0^{\beta_D - 1}C^{\beta_D - 1}
\end{eqnarray}
Again equation A9 is written for $\beta > 1$ and the equivalent form for $\beta < 1$
has no terms in $L_1$ and $\lambda_0^{\beta_D-1}$ is replaced by unity.
Given values for $\beta$ and $\beta_D$, the solution of these equations is elementary.

\begin{thebibliography}{99}
\bibitem{b1} Alexander P., Brown M.T., Scott P.F., 1984, MNRAS, 209, 851
\bibitem{b1a} Alexander P., Pooley G.G., 1996, in Cygnus A -- Study of
a radio galaxy, eds. Carilli C.L., Harris D.E., CUP, p. 149
\bibitem{b2} Alexander P., 2000, MNRAS, 319, 8 
\bibitem{xb2} Baum S.A., Heckman T.M., Bridle A., van Breugel W.J.M.,
Miley G.K., 1988, ApJS, 68, 643
\bibitem{xb2a} Binney J., Tabor G., 1995, MNRAS, 276, 663 
\bibitem{xb2b} Blandford R.D., Rees M.J., 1974, MNRAS, 169, 395
\bibitem{b3} Blanton E.L., Sarazin C.L., McNamara B.R., Wise M.W., 2001,
ApJ, 558, L15 
\bibitem{xb3a} B\"{o}hringer H., Nulsen P.E.J., Braun R., Fabian A.C.,
1995, MNRAS, 274, L67 
\bibitem{xb3b} B\"{o}hringer H., Matsushita K., Churazov E.,
Ikebe Y., Chen Y., 2002 A\&A, 382, 804
\bibitem{b4} Br\"{u}ggen M., Kaiser C.R., 2001, MNRAS, 325, 676 
\bibitem{b5} Carilli C.L., Perley R.A., Harris D.E., 1994, MNRAS, 270, 173
\bibitem{xb5} Churazov E., Br\"{u}ggen M., Kaiser C.R., B\"{o}hringer H., 
Forman W., 2001, 554, 261
\bibitem{b6} Dyson J.E., Falle S.A.E.G., Perry J.J., 1980, MNRAS, 191, 785 
\bibitem{b7} Cowie L.L., McKee C.F., 1977, ApJ, 211, 135 
\bibitem{b8} Ettori S., Fabian A.C., 2000, MNRAS, 317, L57 
\bibitem{b9} Fabian A.C., Sanders J.S., Ettori S., Taylor G.B.,
Allen S.W., Crawford C.S., Iwasawa K., Johnstone R.M., Ogle P.M., 
2000, MNRAS, 318, L65 
\bibitem{b10} Fabian A.C., Celotti A., Blundell K.M., Kassim N.E., Perley R.A., 2002, 
MNRAS, 331, 369 
\bibitem{b10a} David L.P., Nulsen P.E.J., McNamara B.R., Forman W.,
Jones C., Ponman T., Robertson B., WIse M., 2001, ApJ, 557, 546 
\bibitem{b11} Falle S.A.E.G., 1991, MNRAS, 250, 581 
\bibitem{b12} Gull S.G., Northover K.J.E., 1973, Nat., 224, 80 
\bibitem{b13} Hardcastle M.J., Worrall D.M., 2000, MNRAS, 319, 562 
\bibitem{b14} Heinz S., Reynolds C.S., Begelman M.C., 1998, ApJ, 501, 126 (HRB)
\bibitem{b15} Kaiser C.R., Alexander P., 1997, MNRAS 286, 215 (KA97) 
\bibitem{b16} Kaiser C.R., Alexander P., 1999, MNRAS 305, 707 (KA99) 
\bibitem{b17} Leahy J.P., Gizani A.B., 2001, ApJ, 555, 709
\bibitem{xb17a} McNamara B.R., Wise M., Nulsen P.E.J., David L.P.,
Sarazin C.L., Bautz M., Markevitch M., 
Vikhlinin A., Forman W.R., Jones C., Harris D.,
2000a, ApJ, 534, L135
\bibitem{xb17b} McNamara B.R., Wise M.W., David L.P., Nulsen P.E.J.,
Sarazin C.L., 2000b, in  Durret F., Gerbal D. eds., 
Constructing the Universe with Clusters of Galaxies, 
IAP 2000 meeting, Paris, France, contribution reference 6.6
\bibitem{b18a} Nulsen P.E.J., David L.P., McNamara B.R.,
Jones C., Forman W.R., WIse M., 2002, ApJ, 568, 163 
\bibitem{b18} Reynolds C.S., Begelman M.C., 1997, ApJ, 487, L135 
\bibitem{b19} Reynolds C.S., Heinz S., Begelman M.C., 2001, ApJ, 549, L179 
\bibitem{b20} Rizza E., Loken C., Bliton M., Roettiger K., Burns J.O., 2000,
AJ, 119, 21 
\bibitem{xb20a} Saxton C.J., Sutherland R.S., Bicknell G.V., 2002, submitted
(astro-ph/0107558)
\bibitem{b21} Scheuer P.A.G., 1974, MNRAS, 166, 513 
\bibitem{b22} Smith D.A., Wilson A.S., Arnaud K.A., Terashina Y., Young A.J.,
2002, ApJ, in press 
\bibitem{b23} Smith M.D., Smarr L., Norman M.L., Wilson J.R., 1983, ApJ, 264, 432
\bibitem{b24} Yamada M., Fujita Y., 2001, ApJ, 553, L145 
\end{thebibliography}
\end{document}